\documentclass[12pt]{article}

\usepackage{graphics}
\usepackage{epsfig}
\input epsf

\title{Associated production of Higgs boson and heavy quarks at the LHC: predictions with the $k_T$-factorization}

\author{A.V.~Lipatov, N.P.~Zotov}

\begin{document}

\maketitle

\begin{center}

{\it Skobeltsyn Institute of Nuclear Physics,\\ 
Lomonosov Moscow State University,
\\119991 Moscow, Russia\/}\\[3mm]

\end{center}

\vspace{0.5cm}

\begin{center}

{\bf Abstract }

\end{center}

In the framework of the $k_T$-factorization approach, we study the production 
of Higgs bosons associated with a heavy (beauty or top) quark pair at the CERN LHC 
collider conditions. Our consideration is based mainly on the off-shell gluon-gluon fusion
suprocess $g^* g^*  \to Q \bar Q H$. The corresponding matrix element squared 
have been calculated for the first time. 
We investigate the total and differential cross sections of
$b\bar b H$ and $t\bar t H$ production
taking into account also the non-negligible contribution from the $q \bar q \to Q \bar Q H$ mechanism.
In the numerical calculations we use the 
unintegrated gluon distributions obtained from the 
CCFM evolution equation. Our results are compared with the
leading and next-to-leading order predictions of the collinear factorization of QCD.

\vspace{1cm}

\noindent
PACS number(s): 12.38.-t, 12.38.Bx

\vspace{0.5cm}

\section{Introduction} \indent 

It is well known that the electroweak symmetry breaking in the Standard 
Model (SM) of elementary particle interactions is achieved via the Higgs mechanism. 
This mechanism is responsible for the generation of masses of the
gauge ($W^\pm$ and $Z^0$) bosons as well as leptons and quarks via Yukawa couplings.
In the minimal model there are
a single complex Higgs doublet, where the Higgs boson $H$ is the physical neutral
Higgs scalar which is the only remaining part of this doublet after spontaneous
symmetry breaking. In non-minimal models (such as Minimal Supersymmetric Standard Model, MSSM) 
there are additional charged and neutral 
scalar Higgs particles. At moment, the Higgs boson is the only missing, undiscovered
component of modern particle physics, so that
the search for the Higgs boson is of highest priority for particle
physics community. It takes important 
part at the Tevatron experiments and will be one of the main fields of 
study at the LHC collider~[1]. The lower bound on the SM Higgs boson mass
from direct searches at the LEP2 energy is $m_H > 114.4$~GeV~[2], while the
recent global SM fits to electroweak precision data imply $m_H < 211$~GeV~[3].
The MSSM requires the existence of a scalar Higgs boson lighter than about 130~GeV,
so that the possibility of Higgs discovery in the mass range near 115 -- 130~GeV seems
increasingly likely.

The associated production of a Higgs boson with a heavy (beauty or top) quark
pair can play a very important role at high energy hadron colliders.
At the LHC, the $t\bar t H$ production is an important search channel for 
Higgs masses below 130~GeV~[4--6]. Although the expected cross section is rather small,
the signature is quite distinctive. Moreover, analyzing the $t\bar t H$ production
rate can provide information on the top-Higgs Yukawa coupling~[6--9], assuming
standard decay branching ratios~[6], before model independent
precision measurements of this coupling are performed at $e^+ e^-$ colliders~[10--12].
The Higgs boson production in association 
with two beauty quarks is the subject of 
intense theoretical investigations also~[13--15]. In the SM, the coupling of the 
Higgs to a $b \bar b$ pair is suppressed by the small factor $m_b/v$, where
$v = (\sqrt 2 \, G_F)^{-1/2} = 246$~GeV, implying that the $b\bar b H$ production rate
is very small at both the Tevatron and the LHC energies. However, in the MSSM
this coupling grows with the ratio of neutral Higgs boson vacuum expectation
values, $\tan \beta$, and can be significantly enhanced over the SM coupling.
Therefore it is one of the most important discovery channels for supersymmetric
Higgs particles at the LHC.

From the theoretical point of view, the cross section of Higgs and associated heavy quark pair
production at the Tevatron and the LHC is described by the $g g \to Q \bar Q H$ and 
$q \bar q \to Q \bar Q H$ subprocesses (at the tree level). 
The leading-order (LO) QCD predictions~[14, 15] are plagued by considerable uncertainties
due to the strong dependence on the renormalization and factorization scales,
introduced by the QCD coupling and the parton (quark and gluon) densities.
First estimates of radiative corrections were performed~[16] 
in the so-called "effective Higgs approximation" (EHA).
Recently the calculations of the ${\cal O}(\alpha_s^3)$ inclusive 
cross section for the $b\bar b H$ and $t\bar t H$ production have been 
carried out at next-to-leading order (NLO)~[17--20] of QCD. 
These calculations are based on the complete set of virtual and real 
${\cal O}(\alpha_s)$ corrections to the parton level processes $g g \to Q \bar Q H$ and
$qq \to Q\bar Q H$, as well as the tree level process $(q,\bar q) g \to Q \bar Q H + (q,\bar q)$.
The NLO cross sections are about 20\% smaller and about 30\% larger than the relevant LO cross 
sections at the Tevatron and LHC conditions, respectively. 
It was demonstrated~[19, 20] that these high-order QCD corrections greatly reduce the 
renormalization and factorization scale dependence of LO results and thus stabilize the 
theoretical predictions.

In the present paper we will study the Higgs and associated heavy (beauty and top) quark pair
production using the so-called $k_T$-factorization QCD approach~[21--24]. This approach 
is based on the familiar Balitsky-Fadin-Kuraev-Lipatov (BFKL)~[25] or 
Ciafaloni-Catani-Fiorani-Marchesini 
(CCFM)~[26] equations for the non-collinear gluon evolution in a proton.
%The basic dynamical quantity of the $k_T$-factorization approach is 
%the so-called unintegrated (i.e. ${\mathbf k}_T$-dependent) gluon distribution 
%$f_g(x,{\mathbf k}_T^2,\mu^2)$ which determines the probability to find a 
%gluon carrying the longitudinal momentum fraction $x$ and the transverse momentum 
%${\mathbf k}_T$ at the probing scale $\mu^2$.
Detailed description of the $k_T$-factorization approach can be
found, for example, in reviews~[27--29].
Here we would like to only mention that
the main part of high-order radiative QCD corrections is naturally included
into the leading-order $k_T$-factorization formalism.

The $k_T$-factorization approach has been already applied~[30--35] to study the 
inclusive Higgs production at the Tevatron and LHC conditions.
First investigations~[30--33] were based on the 
amplitude for scalar Higgs boson production in the fusion of
two off-shell gluons $g^* g^* \to H$. The corresponding matrix elements have been 
derived first in~[36] using the large $m_t$ limit where the effective Lagrangian~[37] for the Higgs 
boson coupling to gluons can be applied\footnote{The calculations~[30, 32] were performed 
using the relevant on-mass shell matrix element.}. 
The investigations~[30--33] provocated further studies~[34, 35] where 
the off-shell matrix elements of $g^* g^* \to H$ subprocess have 
been calculated including finite masses of quarks in the triange loop.
It was claimed~[35] that the $k_T$-factorization approach give us the possibility
to estimate the size of unknown collinear high-order corrections.

The starting point of present consideration is the off-shell amplitude of 
gluon-gluon fusion suprocess $g^* g^* \to Q \bar Q H$. We evaluate
the corresponding matrix elements squared for the first time
and apply them for investigation of the $b\bar b H$ and $t\bar t H$ production
rates at the LHC energy, $\sqrt s = 14$~TeV. The quark-antiquark annihilation mechanism, 
$q\bar q \to Q \bar Q H$,
is expected to be significant only at relatively large $x$, and therefore
we can safely take it into accout in the usual leading-order collinear approximation of QCD.
In the numerical calculations we will use the unintegrated gluon density in a proton 
which was obtained~[38] from the CCFM equation.
Of course, we expect that effects coming from the non-zero gluon virtualities for 
the associated $b\bar b H$ and $t \bar t H$ production are not very well prononced even at LHC energies.
% because relatively large Higgs mass, $m_H \sim 120$~GeV.
However, our study is important since 
it is planned to include
the calculated off-shell matrix element $g^* g^* \to Q \bar Q H$ 
to the Monte-Carlo generator \textsc{Cascade}~[39].
We will compare the results obtained in the $k_T$-factorization approach with the
leading and next-to-leading order predictions of the collinear factorization of QCD.

The outline of our paper is following. In Section~2 we 
recall shortly the basic formulas of the $k_T$-factorization approach with a brief 
review of calculation steps. 
We will concentrate mainly on the $g^* g^* \to Q \bar Q H$
subprocess. The evaluation of $q \bar q\to Q \bar Q H$ 
contribution is a rather straightforward and therefore will not
discussed here (for the reader's convenience, we only collect 
the relevant formulas in Appendix).
In Section~3 we present the numerical results
of our calculations and a discussion. 
Section~4 contains our conclusions.

\section{Theoretical framework} 
\subsection{Kinematics} \indent 

We start from the kinematics (see Fig.~1). 
Let $p^{(1)}$ and $p^{(2)}$ be the four-momenta of the incoming protons and 
$p$ be the four-momentum of the produced Higgs boson.
The initial off-shell gluons have the four-momenta
$k_1$ and $k_2$ and the final quark and antiquark have the 
four-momenta $p_1$ and $p_2$ and masses $m_Q$, respectively.
In the proton-proton center-of-mass frame we can write
$$
  p^{(1)} = {\sqrt s}/2\,(1,0,0,1),\quad p^{(2)} = {\sqrt s}/2\,(1,0,0,-1), \eqno(1)
$$

\noindent
where $\sqrt s$ is the total energy of the process 
under consideration and we neglect the masses of the incoming protons.
The initial gluon four-momenta in the high energy limit can be written as
$$
  k_1 = x_1 p^{(1)} + k_{1T},\quad k_2 = x_2 p^{(2)} + k_{2T}, \eqno(2)
$$

\noindent 
where $k_{1T}$ and $k_{2T}$ are their transverse four-momenta.
It is important that ${\mathbf k}_{1T}^2 = - k_{1T}^2 \neq 0$ and
${\mathbf k}_{2T}^2 = - k_{2T}^2 \neq 0$. From the conservation laws 
we can easily obtain the following conditions:
$$
  {\mathbf k}_{1T} + {\mathbf k}_{2T} = {\mathbf p}_{1T} + {\mathbf p}_{2T} + {\mathbf p}_{T},
$$
$$
  x_1 \sqrt s = m_{1T} e^{y_1} + m_{2T} e^{y_2} + m_{T} e^y, \eqno(3)
$$
$$
  x_2 \sqrt s = m_{1T} e^{-y_1} + m_{2T} e^{-y_2} + m_{T} e^{-y},
$$

\noindent 
where $y$ and $m_{T}$ are the rapidity and the transverse mass of the produced Higgs boson, 
$p_{1T}$ and $p_{2T}$ are the transverse four-momenta of the final quark and antiquark, 
$y_1$, $y_2$, $m_{1T}$ and $m_{2T}$ are their center-of-mass rapidities and 
transverse masses, i.e. $m_{iT}^2 = m_Q^2 + {\mathbf p}_{iT}^2$.

\subsection{Off-shell amplitude of the $g^* g^* \to Q \bar Q H$ subprocess} \indent 

There are eight Feynman diagrams (see Fig.~2) which describe the partonic
subprocess $g^* g^* \to Q \bar Q H$ at $\alpha \alpha_s^2$ order.
Let $\epsilon_1$ and $\epsilon_2$ be the initial off-shell gluon 
polarization vectors and $a$ and $b$ the relevant eight-fold color indices.
Then the relevant matrix element can be presented as follows:
$$
  {\cal M}_1 = g^2 \, \bar u (p_1) \, t^a \gamma^\mu \epsilon_\mu {\hat p_1 - \hat k_1 + m_1\over m_1^2 - (p_1 - k_1)^2} H {\hat k_2 - \hat p_2 + m_2\over m_2^2 - (k_2 - p_2)^2} t^b \gamma^\nu \epsilon_\nu \, u(p_2), \eqno(4)
$$
$$
  {\cal M}_2 = g^2 \, \bar u (p_1) \, t^b \gamma^\nu \epsilon_\nu {\hat p_1 - \hat k_2 + m_1\over m_1^2 - (p_1 - k_2)^2} H {\hat k_1 - \hat p_2 + m_2\over m_2^2 - (k_1 - p_2)^2} t^a \gamma^\mu \epsilon_\mu \, u(p_2), \eqno(5)
$$
$$
  {\cal M}_3 = g^2 \, \bar u (p_1) \, t^a \gamma^\mu \epsilon_\mu {\hat p_1 - \hat k_1 + m_1\over m_1^2 - (p_1 - k_1)^2}\, t^b \gamma^\nu \epsilon_\nu { - \hat p_2 - \hat p + m_1\over m_1^2 - ( - p_2 - p)^2} H \, u(p_2), \eqno(6)
$$
$$
  {\cal M}_4 = g^2 \, \bar u (p_1) \, t^b \gamma^\nu \epsilon_\nu {\hat p_1 - \hat k_2 + m_1\over m_1^2 - (p_1 - k_2)^2}\, t^a \gamma^\mu \epsilon_\mu { - \hat p_2 - \hat p + m_1\over m_1^2 - ( - p_2 - p)^2} H \, u(p_2), \eqno(7)
$$
$$
  {\cal M}_5 = g^2 \, \bar u (p_1) \, H {\hat p_1 + \hat p + m_2\over m_2^2 - (p_1 + p)^2}\, t^b \gamma^\nu \epsilon_\nu { \hat k_1 - \hat p_2 + m_2\over m_2^2 - (k_1 - p_2)^2} t^a \gamma^\mu \epsilon_\mu \, u(p_2), \eqno(8)
$$
$$
  {\cal M}_6 = g^2 \, \bar u (p_1) \, H {\hat p_1 + \hat p + m_2\over m_2^2 - (p_1 + p)^2}\, t^a \gamma^\mu \epsilon_\mu { \hat k_2 - \hat p_2 + m_2\over m_2^2 - (k_2 - p_2)^2} t^b \gamma^\nu \epsilon_\nu \, u(p_2), \eqno(9)
$$
$$
  \displaystyle {\cal M}_7 = g^2 \, \bar u (p_1) \, \gamma^\rho C^{\mu \nu \rho}(k_1,k_2,- k_1 - k_2){\epsilon_\mu \epsilon_\nu \over (k_1 + k_2)^2} f^{abc} t^c \times \atop 
  \displaystyle \times { - \hat p_2 - \hat p + m_1\over m_1^2 - ( - p_2 - p)^2}\, H \, u(p_2), \eqno(10)
$$
$$
  \displaystyle {\cal M}_8 = g^2 \, \bar u (p_1) \, H \, \epsilon_\lambda {\hat p_1 + \hat p + m_2\over m_2^2 - (p_1 + p)^2} \times \atop 
  \displaystyle \times \gamma^\rho C^{\mu \nu \rho}(k_1,k_2,- k_1 - k_2) {\epsilon_\mu \epsilon_\nu \over (k_1 + k_2)^2} f^{abc} t^c \, u(p_2). \eqno(11)
$$

\vspace{0.2cm}

\noindent
In the above expressions $C^{\mu \nu \rho}(k,p,q)$ and $H$ are related to the standard QCD
three-gluon coupling and the $H$-fermion vertexes:
$$
  C^{\mu \nu \rho}(k,p,q) = g^{\mu \nu} (p - k)^\rho + g^{\nu \rho} (q - p)^\mu + g^{\rho \mu} (k - q)^\nu, \eqno(12)
$$
$$
  H = - {e\over \sin 2 \theta_W} {m_Q\over m_Z}, \eqno(14)
$$

\noindent
where $\theta_W$ is the Weinberg mixing angle and $m_Z$ is the $Z$-boson mass.
The summation on the initial off-shell gluon polarizations is carried out using the
BFKL prescription~[21--25]:
$$
  \sum \epsilon^\mu (k_i) \, \epsilon^{  \, \nu} (k_i) = {k_{iT}^\mu k_{iT}^\nu \over {\mathbf k}_{iT}^2}. \eqno(15)
$$

\noindent
This formula converges to the usual expression 
$\sum \epsilon^\mu \epsilon^{  \, \nu} = -g^{\mu \nu}$ 
after azimuthal angle averaging in the $k_T \to 0$ limit. 
The evaluation of the traces in~(4) --- (11) was done using the algebraic 
manipulation system \textsc{Form}~[40]. 
We would like to mention here that the usual method 
of squaring of~(4) --- (11) results in enormously long
output. This technical problem was solved by applying the
method of orthogonal amplitudes~[41].

The gauge invariance of the matrix element is a
subject of special attention in the $k_T$-factorization approach. Strictly speaking,
the diagrams shown in Fig.~2 are insufficient and have to be accompanied
with the graphs involving direct gluon exchange between the protons
(these protons are not shown in Fig.~2). These graphs are 
necessary to maintain the gauge invariance.
However, they violate the factorization since they cannot be represented
as a convolution of the gluon-gluon fusion matrix element with unintegrated gluon density.
The solution pointed out in~[23, 24] refers to the fact that, within the 
particular gauge~(15), the contribution from these unfactorizable diagrams
vanish, and one has to only take into account the graphs depicted in Fig.~2.
We have successfully tested the gauge invariance of the matrix 
element~(4) --- (11) numerically.

\subsection{Cross section for the $Q \bar Q H$ production} \indent 

According to the $k_T$-factorization theorem, the 
$Q \bar Q H$ production cross section 
via two off-shell gluon fusion can be written as a convolution
$$
  \displaystyle \sigma (p p \to Q \bar Q H) = \int {dx_1\over x_1} f_g(x_1,{\mathbf k}_{1 T}^2,\mu^2) d{\mathbf k}_{1 T}^2 {d\phi_1\over 2\pi} \times \atop 
  \displaystyle \times \int {dx_2\over x_2} f_g(x_2,{\mathbf k}_{2 T}^2,\mu^2) d{\mathbf k}_{2 T}^2 {d\phi_2\over 2\pi} d{\hat \sigma} (g^* g^* \to Q \bar Q H), \eqno(16)
$$

\noindent 
where $\hat \sigma(g^* g^* \to Q \bar Q H)$ is the partonic cross section, 
$f_g(x,{\mathbf k}_{T}^2,\mu^2)$ is the unintegrated gluon distribution in a proton 
and $\phi_1$ and $\phi_2$ are the azimuthal angles of the incoming gluons.
The multiparticle phase space $\Pi d^3 p_i / 2 E_i \delta^{(4)} (\sum p^{\rm in} - \sum p^{\rm out} )$
is parametrized in terms of transverse momenta, rapidities and azimuthal angles:
$$
  { d^3 p_i \over 2 E_i} = {\pi \over 2} \, d {\mathbf p}_{iT}^2 \, dy_i \, { d \phi_i \over 2 \pi}. \eqno(17)
$$

\noindent
Using the expressions~(16) and~(17) we obtain the master formula:
$$
  \displaystyle \sigma(p p \to Q\bar Q H) = \int {1\over 256\pi^3 (x_1 x_2 s)^2} |\bar {\cal M}(g^* g^* \to Q\bar Q H)|^2 \times \atop 
  \displaystyle \times f_g(x_1,{\mathbf k}_{1 T}^2,\mu^2) f_g(x_2,{\mathbf k}_{2 T}^2,\mu^2) d{\mathbf k}_{1 T}^2 d{\mathbf k}_{2 T}^2 d{\mathbf p}_{1 T}^2 {\mathbf p}_{2 T}^2 dy dy_1 dy_2 {d\phi_1\over 2\pi} {d\phi_2\over 2\pi} {d\psi_1\over 2\pi} {d\psi_2\over 2\pi}, \eqno(18)
$$

\noindent
where $|\bar {\cal M}(g^* g^* \to Q \bar Q H)|^2$ is the off-mass shell 
matrix element squared and averaged over initial gluon 
polarizations and colors, $\psi_1$ and $\psi_2$ are the 
azimuthal angles of the final state quark and antiquark, respectively.
We would like to point out again that $|\bar {\cal M}(g^* g^* \to Q \bar Q H)|^2$
strongly depends on the nonzero 
transverse momenta ${\mathbf k}_{1 T}^2$ and ${\mathbf k}_{2 T}^2$.
If we average the expression~(18) over $\phi_{1}$ and $\phi_{2}$ 
and take the limit ${\mathbf k}_{1 T}^2 \to 0$ and ${\mathbf k}_{2 T}^2 \to 0$,
then we recover the expression for the $Q\bar Q H$ production cross section in the  
collinear $\alpha \alpha_s^2$ approximation.

The multidimensional integration in~(18) has been performed
by means of the Monte Carlo technique, using the routine 
\textsc{Vegas}~[42]. The full C$++$ code is available from the 
authors upon request\footnote{lipatov@theory.sinp.msu.ru}.

\section{Numerical results} \indent

We now are in a position to present our results. 
According to (18), in the numerical calculations below
we have used the CCFM-evolved unintegrated gluon density in a proton, namely set $A0$~[38]. This
set is widely discussed in the literature\footnote{See, for example, review~[29] for
more information.} and has been implemented in the Monte-Carlo generator \textsc{Cascade}~[39]. As it often done for $Q\bar Q H$ production~[17--20], we choose the renormalization and 
factorization scales to be equal: $\mu_R = \mu_F =  \xi (m_Q + m_H/2)$. 
In order to investigate the scale dependence of our 
results we will vary the scale parameter
$\xi$ between $1/2$ and~2 about the default value $\xi = 1$.
As it was proposed in~[38], for $\xi = 1/2$ and $\xi = 2$ we use the $A0-$ and 
$A0+$ sets of unintegrated gluon densities, respectively.
For completeness, we set to
$m_b = 4.75$~GeV, $m_t = 172$~GeV, $m_Z = 91.1876$~GeV,
$\sin^2 \theta_W = 0.23122$ and use the LO formula for the 
coupling constant $\alpha_s(\mu^2)$ 
with $n_f = 4$ active quark flavours at $\Lambda_{\rm QCD} = 200$~MeV, such 
that $\alpha_s(M_Z^2) = 0.1232$.

We begin the discussion by presenting our numerical results
for the associated $b\bar b H$ and $t\bar t H$ total cross sections
as a function of Higgs boson mass for the LHC energy, $\sqrt s = 14$~TeV.
We consider $100 < m_H < 200$~GeV since the production of a Higgs
boson in association with a pair of beauty or top quarks at the LHC 
will play an important role only for relatively light Higgs bosons.
The solid histograms in Fig.~3 correspond to the results obtained in 
the $k_T$-factorization approach of QCD with the CCFM-evolved gluon density.
The theoretical uncertainties of these predictions are presented by 
upper and lower dashed histograms.
The dash-dotted histograms represent results which were obtained 
in the standard (collinear) approximation of QCD at LO\footnote
{Numerically, we have used the standard GRV~(LO) parametrizations~[43] of 
collinear parton densities.}.
The contributions from the quark-antiquark annihilation mechanism,
$q\bar q \to Q\bar Q H$, are shown by the dotted histograms.
One can see that using of the $k_T$-factorization approach
leads to some enhancement of the predicted $b\bar b H$ cross section at 
low $m_H$ region (namely $m_H < 150$~GeV) in respect to the collinear LO QCD results. 
In the case of $t\bar t H$
production, the calculated cross sections in both approaches are very close to each other.
It is because the large-$x$ region, namely $x \sim 0.1$, is only covered here and
therefore there is practically no effects connected with the small-$x$ physics.
It is important that we find our leading-order predictions fully consistent with the corresponding 
LO results presented in~[17--20]. The small visible differense can easily come from the 
different quark and gluon densities\footnote{The LO parton densities 
from CTEQ5L set~[44] have been used in~[17--20].}.

In contrast with the NLO QCD results, the $k_T$-factorization
approach not reduces the strong scale dependence of corresponding LO QCD predictions, 
which has been pointed out in~[17--20]. 
We conservatively estimate this theoretical uncertainty to be at
most of order $40 - 50$\% (see Fig.~3). Such scale dependence is significant, of course, and this 
fact indicates the necessarity of inclusion of the high-order corrections to the
$k_T$-factorization formalism. So far the $k_T$-factorization is 
based on the leading-order BFKL or CCFM evolution equations. 
On the other hand, the kernel of BFKL equation has been calculated already 
at NLO~[45],
so that in the small-$x$ regime the $k_T$-factorization can be formulated at 
NLO accuracy also~[46]. At moment, this problem is not solved and much more 
further efforts should be concentrated in this field.
We only mention here that the leading-order $k_T$-factorization naturally includes
the high-energy part of the NLO collinear corrections.

Our predictions for the transverse momentum and rapidity distributions of the Higgs boson 
as well as associated beauty or top quark $Q$ are shown in Figs.~4 --- 7.
The distributions on the azimuthal angle distance between the $H$ and $Q$ as well as 
the $Q$ and $\bar Q$ are shown in Figs.~8 and 9, respectively.
These calculations were performed using $m_H = 120$~GeV.
The comparison of the $k_T$-factorization approach to the collinear one shows
the some broadening of the transverse momentum distributions 
due to extra transverse momentum of the colliding off-shell gluons.
Also the $k_T$-factorization result shows a more homogeneous spread
of the azimuthal angle $\Delta \phi^{H - b}$ distance.
At the same time, the cross sections calculated as a function of rapidities $y^H$ and $y^Q$ 
as well as the azimuthal angle distributions $d\sigma/d\Delta \phi^{Q - \bar Q}$
show a similar behaviour, except for the overall normalization.

To elaborate the difference between the $k_T$-factorization approach
and the collinear approximation of QCD, we investigate more exclusive
observables, like the cross section differential in the total transverse
momentum of the $Q\bar Q H$ system, $p_T^{Q\bar Q H}$.
In the usual collinear factorization of QCD the effect of
intrinsic transverse momenta of the initial gluons can not be described until
higher order corrections are taken into account. In the NLO QCD a
non-zero $p_T^{Q\bar Q H}$ is generated by the emission of an additional
gluon, while at LO it is always balanced to zero.
In the $k_T$-factorization formalism, taking into account the non-vanishing 
initial gluon transverse momentum ${\mathbf k}_{T}$ leads to 
the violation of back-to-back kinematics even at leading order.
This effect is clearly illustrated in Fig.~10, where we plot the
$b\bar b H$ and $t\bar t H$ cross sections as a function of $p_T^{Q\bar Q H}$.
Note that only the off-shell gluon-gluon mechanism, 
$g^* g^* \to Q\bar Q H$, has been taken into account here. 
The relevant contribution from the quark-antiquark annihilation,
$q \bar q\to Q\bar Q H$, is expected to be almost negligible for $b\bar b H$ production
and probably can be sizeble for $t\bar t H$ one\footnote{We do not consider here the problem of proper 
transverse momentum generation of initial state quarks.}.
Keeping in mind that the NLO for this observable is the first non-trivial
order, it would be useful to compare the NLO QCD and $k_T$-factorization predictions
in order to investigate the exact effect of high-order corrections in collinear factorization.

In addition, we evaluate the fully exclusive cross section for $b\bar b H$ production
by requiring that the transverse momentum of one or both final state beauty quarks be 
large than some $p_T^{\rm cut}$ value. This corresponds to an experiment 
measuring the Higgs decay products along with one or two high $p_T$ beauty quark jets
that are clearly separated from the beam. In Fig.~11 we illustrate the dependence
of these exclusive cross sections on the $p_T^{\rm cut}$ parameter.
Reducting the $p_T^{\rm cut}$ from 50~GeV to zero approximately increases
the relevant cross sections by a factors of about 10 and 100, respectively.
In the collinear factorization of QCD, if both beauty quarks are required to be produced with
$p_T > 20$~GeV, the NLO corrections reduce the LO predictions, and these corrections are
positive if beauty quarks produced at small $p_T$~[17]. 
However, one can see that predictions of the $k_T$-factorization approach overestimate 
the collinear LO results in a wide $p_T^{\rm cut}$ range.

Finally, we would like to mention that our $k_T$-factorization calculations can be
straightforwardly generalized to the case of scalar Higgs bosons of the MSSM
by replacing the SM beauty and top quark Yukawa couplings with the
corresponding MSSM ones. It is because the off-mass shell matrix element
calculated above (see Section~2.2) is proportional to the beauty and top quark Yukawa couplings.
In the MSSM, these couplings to the scalar Higgs bosons, $\hat g_{QQ H}$, are given by a simple rescaling of SM 
couplings $g_{QQ H}$~[47], i.e. 

$$ 
  \displaystyle \hat g_{bbh^0} = - {\sin \alpha\over \cos \beta } \, g_{bbh},\quad \hat g_{tth^0} = {\cos \alpha\over \sin \beta } \, g_{tth} \atop
  \displaystyle \hat g_{bbH^0} = {\cos \alpha\over \cos \beta } \, g_{bbh},\quad \hat g_{ttH^0} = {\sin \alpha\over \sin \beta } \, g_{tth}, \eqno(19)
$$

\noindent
where $h^0$ and $H^0$ are the lighter and heavier neutral scalars of MSSM, and $\alpha$ is the
angle which diagonalizes the neutral scalar Higgs mass matrix.
However, at the NLO level this rescaling is spoiled by one-loop diagrams in which the Higgs boson couples 
to a closed quark loop. We do not consider supersymmetric-QCD corrections in this paper.

\section{Conclusions} \indent 

We have studied the associated production of Higgs boson and
beauty or top quark pair in hadronic collisions at the LHC conditions
in the $k_T$-factorization approach of QCD.
Our consideration is based on the amplitude of off-shell gluon-gluon 
fusion subprocess $g^*g^* \to Q \bar{Q} H$. The corresponding off-shell 
matrix elements have been calculated for the first time.
Sizeble contributions from the $q \bar q \to Q \bar Q H$ mechanism have been
taken into account in the LO approximation of collinear QCD.

We have investigated the total and differential cross sections of
$b\bar b H$ and $t\bar t H$ production. 
In the numerical calculations we have used the 
unintegrated gluon distributions obtained from the 
CCFM evolution equation. The comparisons with the
leading and next-to-leading order QCD predictions have been made.
We demonstrate that the $k_T$-factorization
approach not reduces the strong scale dependence of collinear LO QCD predictions,
pointed out in~[17--20].
This fact indicates the importance of the high-order correction within the 
$k_T$-factorization approach. These corrections should be developed and 
taken into account in the future applications.
Finally, we show how our results can be generalized to the scalar 
Higgs sector of the MSSM. Our calculations is also important for Higgs boson searches where
one or two high-$p_T$ beauty quarks are tagged in final state.

\section{Acknowledgements} \indent 

We thank S.P.~Baranov for the cross-check of matrix elements
and very helpful discussions, H.~Jung for his encouraging interest
and for providing the CCFM code for 
unintegrated gluon distributions. 
The authors are very grateful to 
DESY Directorate for the support in the 
framework of Moscow --- DESY project on Monte-Carlo
implementation for HERA --- LHC.
A.V.L. was supported in part by the grants of the president of 
Russian Federation (MK-438.2008.2) and Helmholtz --- Russia
Joint Research Group.
Also this research was supported by the 
FASI of Russian Federation (grant NS-8122.2006.2)
and the RFBR fundation (grant 08-02-00896-a).

\section{Appendix A} \indent 

Here we present the compact analytic expressions for the 
$q\bar q \to Q\bar Q H$ subprocess.
Let us define the four-momenta of the incoming and outgoing quark as 
$k_1$, $k_2$, $p_1$ and $p_2$, respectively. 
The outgoing quarks have mass $m_Q$, i.e. $p_1^2 = p_2^2 = m_Q^2$.
In the formulas below we will neglect the
masses of the incoming quarks.

The contribution of the $q\bar q \to Q\bar Q H$ subrocess to the total $Q\bar Q H$ cross section 
can be easily calculated using the master formula~(18). 
One should only replace the unintegrated gluon
densities $f_g(x,{\mathbf k}_{T}^2,\mu^2)$ by the quark ones,
perform the summation over initial quark flavours and take the collinear limit.
The squared leading-order matrix elements $|\bar {\cal M}(q \bar q \to Q \bar Q H)|^2$ 
summed over final polarization states and averaged over initial ones can be written as follows:
$$
  |\bar {\cal M}(q \bar q \to Q \bar Q H)|^2 = { (4 \pi)^3 \over 72 \sin^2 2 \theta_W} \left({m_Q \over m_Z}\right)^2 \alpha \alpha_s^2 \, \left[ { F_{11}\over T_1^2} + { F_{22}\over T_2^2} + { F_{12} + F_{21}\over T_1 T_2}\right], \eqno(A.1)
$$

\noindent 
where
$$
  F_{11} = 128 \,(p_1 p_2) (p_2 k_1) (p_2 k_2) - 64 \, (p_1 p_2) (p_2 k_1) (k_1 k_2) - 64 \, (p_1 p_2) (p_2 k_2) (k_1 k_2)  - 
$$
$$
  64 \, (p_1 p_2) (k_1 k_2) m_Q^2 - 64 \, (p_1 k_1) (p_2 k_1) (p_2 k_2) + 64 \, (p_1 k_1) (p_2 k_1) (k_1 k_2) + 
$$
$$
  64 \, (p_1 k_1) (p_2 k_2)^2 + 64 \, (p_1 k_1) (k_1 k_2) m_Q^2 - 64 \, (p_1 k_2) (p_2 k_1) (p_2 k_2) + 
$$
$$
  64 \, (p_1 k_2) (p_2 k_1)^2 + 64 \, (p_1 k_2) (p_2 k_2) (k_1 k_2) + 64 \, (p_1 k_2) (k_1 k_2) m_Q^2 - 
$$
$$
  128 \, (p_2 k_1) (p_2 k_2) m_Q^2 + 64 \, (k_1 k_2) m_Q^4 + 64 \, (k_1 k_2)^2 m_Q^2, \eqno(A.2)
$$
$$
  F_{22} = 128 \, (p_1 p_2) (p_1 k_1) (p_1 k_2) - 64 \, (p_1 p_2) (p_1 k_1) (k_1 k_2) - 64 \, (p_1 p_2) (p_1 k_2) (k_1 k_2) - 
$$
$$
  64 \, (p_1 p_2) (k_1 k_2) m_Q^2 - 64 \, (p_1 k_1) (p_1 k_2) (p_2 k_1) - 64 \, (p_1 k_1) (p_1 k_2) (p_2 k_2) - 
$$
$$
  128 \, (p_1 k_1) (p_1 k_2) m_Q^2 + 64 \, (p_1 k_1) (p_2 k_1) (k_1 k_2) + 64 \, (p_1 k_1)^2 (p_2 k_2) + 
$$
$$
  64 \, (p_1 k_2) (p_2 k_2) (k_1 k_2) + 64 \, (p_1 k_2)^2 (p_2 k_1) + 64 \, (p_2 k_1) (k_1 k_2) m_Q^2 + 
$$
$$
  64 \, (p_2 k_2) (k_1 k_2) m_Q^2 + 64 \, (k_1 k_2) m_Q^4 + 64\, (k_1 k_2)^2 m_Q^2, \eqno(A.3)
$$
$$
  F_{12} = F_{21} = - 64 \, (p_1 p_2) (p_1 k_1) (p_2 k_2) - 32 \, (p_1 p_2) (p_1 k_1) (k_1 k_2) - 64 \, (p_1 p_2) (p_1 k_2) (p_2 k_1) - 
$$
$$
  32 \, (p_1 p_2) (p_1 k_2) (k_1 k_2) - 32 \, (p_1 p_2) (p_2 k_1) (k_1 k_2) - 32 \, (p_1 p_2) (p_2 k_2) (k_1 k_2) - 
$$ 
$$
  64 \, (p_1 p_2) (k_1 k_2) m_Q^2 + 64 \, (p_1 p_2)^2 (k_1 k_2) - 32 \, (p_1 k_1) (p_1 k_2) (p_2 k_1) - 
$$ 
$$
  32 \, (p_1 k_1) (p_1 k_2) (p_2 k_2) - 32 \, (p_1 k_1) (p_2 k_1) (p_2 k_2)  + 64 \, (p_1 k_1) (p_2 k_1) (k_1 k_2) + 
$$
$$
  64 \, (p_1 k_1) (p_2 k_2) m_Q^2 + 32 \, (p_1 k_1) (p_2 k_2)^2 + 32 \, (p_1 k_1) (k_1 k_2) m_Q^2 + 
$$
$$ 
  32 \, (p_1 k_1)^2 (p_2 k_2) - 32 \, (p_1 k_2) (p_2 k_1) (p_2 k_2) + 64 \, (p_1 k_2) (p_2 k_1) m_Q^2 + 
$$
$$
  32 \, (p_1 k_2) (p_2 k_1)^2 + 64 \, (p_1 k_2) (p_2 k_2) (k_1 k_2) + 32 \, (p_1 k_2) (k_1 k_2) m_Q^2 + 
$$
$$
  32 \, (p_1 k_2)^2 (p_2 k_1) + 32 \, (p_2 k_1) (k_1 k_2) m_Q^2 + 32 \, (p_2 k_2) (k_1 k_2) m_Q^2 + 64 \, (k_1 k_2)^2 m_Q^2, \eqno(A.4)
$$

\noindent
$$
  T_1 = (p_2 k_1) + (p_2 k_2) - (k_1 k_2), \quad T_2 =  (p_1 k_1) + (p_1 k_2) - (k_1 k_2). \eqno(A.5)
$$

\newpage

\begin{figure}
\begin{center}
\epsfig{figure=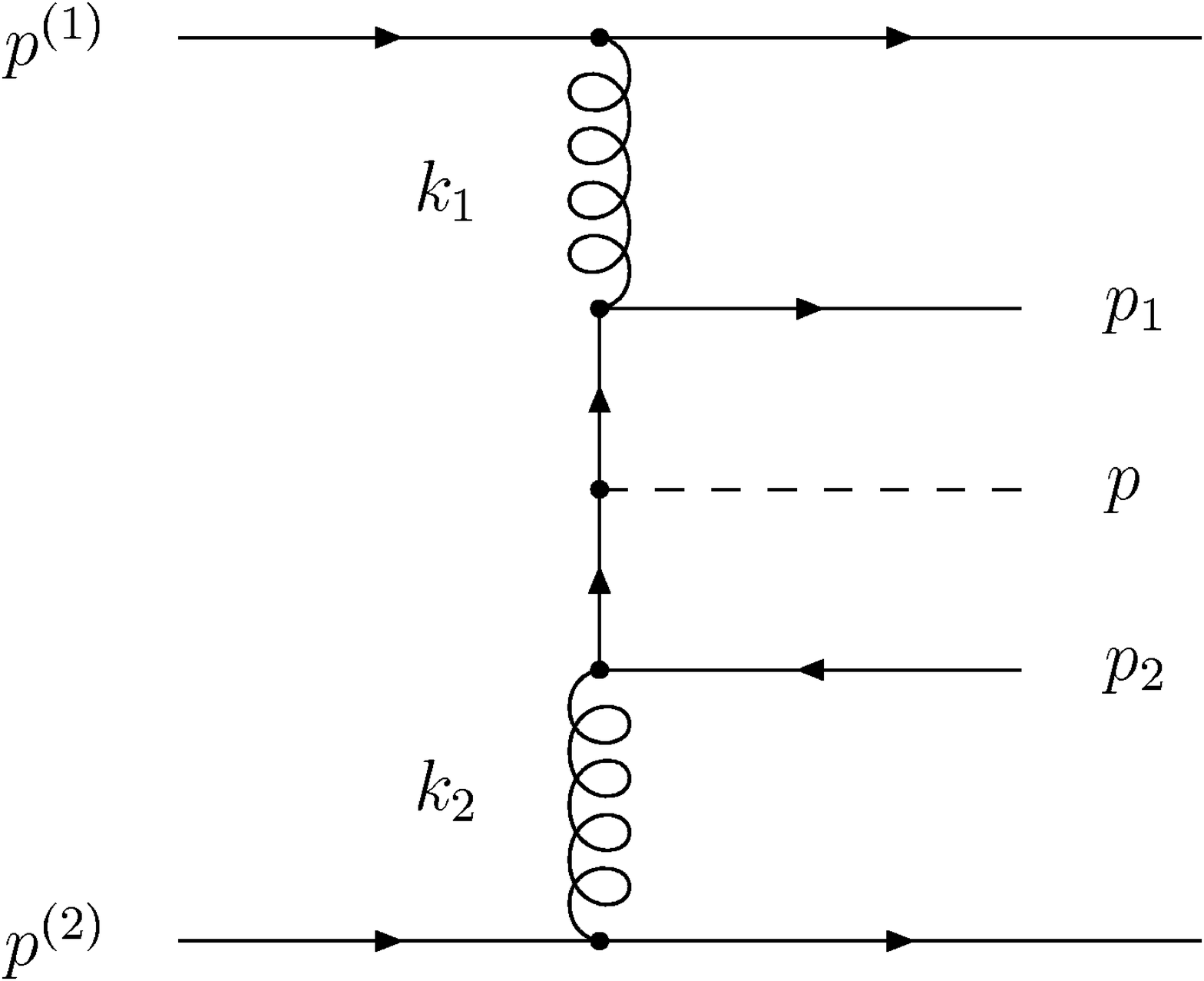, width = 10cm}
\caption{Kinematics of the $g^* g^* \to Q \bar Q H$ process.}
\label{fig1}
\end{center}
\end{figure}

\newpage

\begin{figure}
\begin{center}
\epsfig{figure=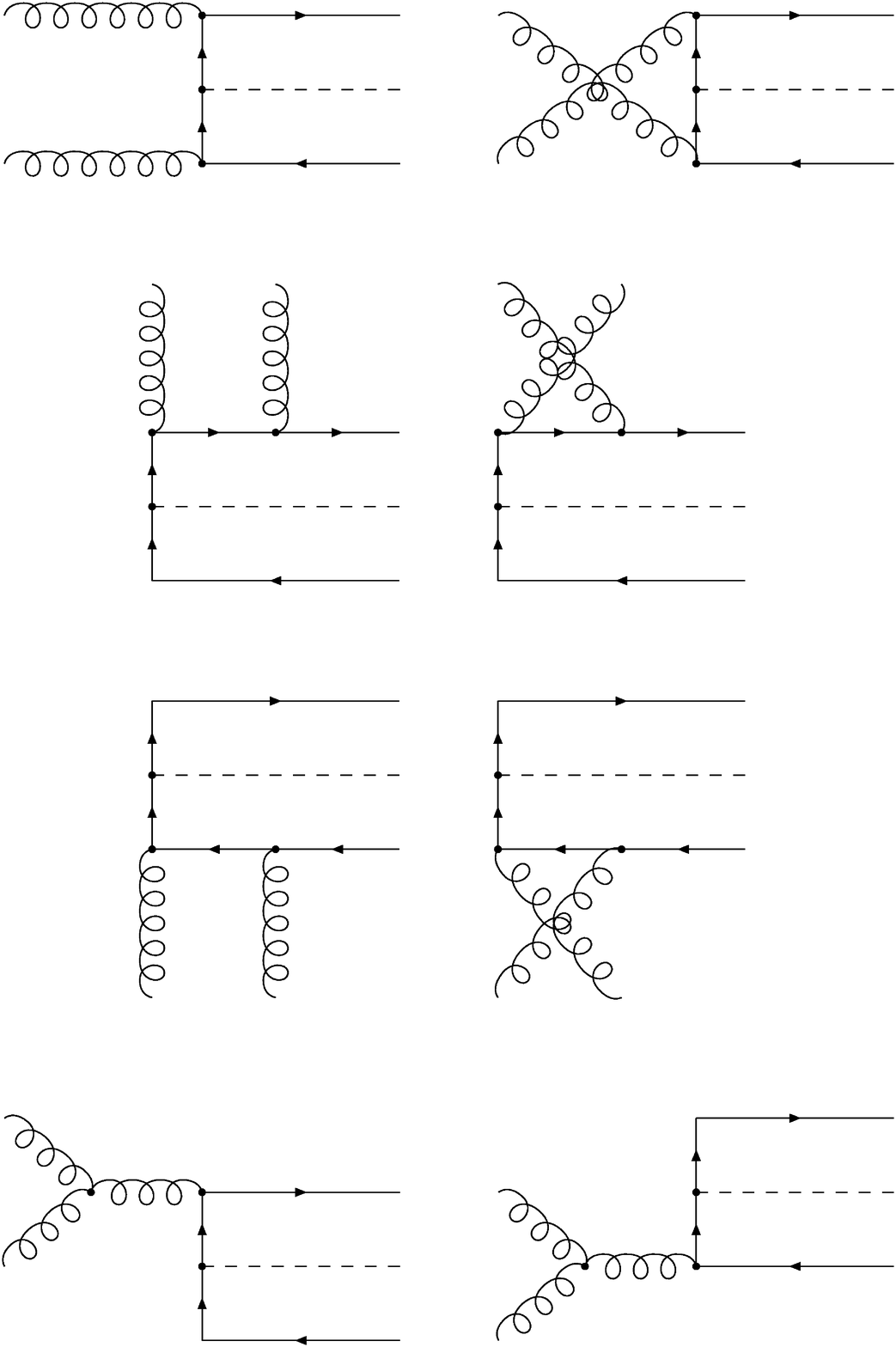, width = 14cm}
\caption{Feynman diagrams which describe the partonic
subprocess $g^* + g^* \to Q \bar Q H$ at the leading order in $\alpha_s$ and $\alpha$.}
\label{fig2}
\end{center}
\end{figure}

\newpage

\begin{figure}
\begin{center}
\epsfig{figure=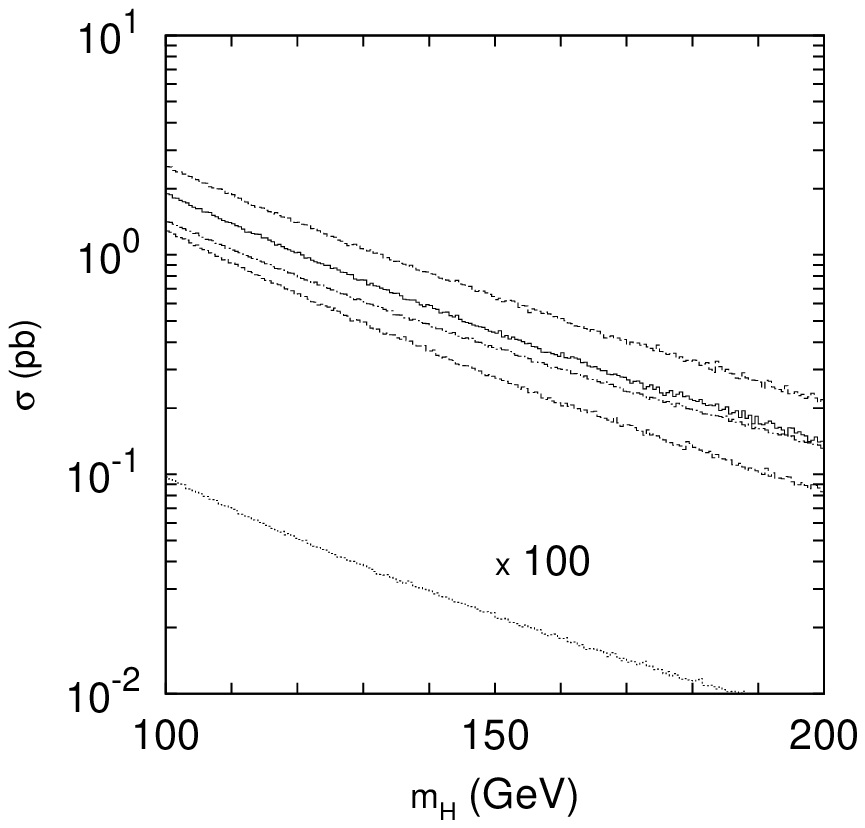, width = 13.0cm}
\epsfig{figure=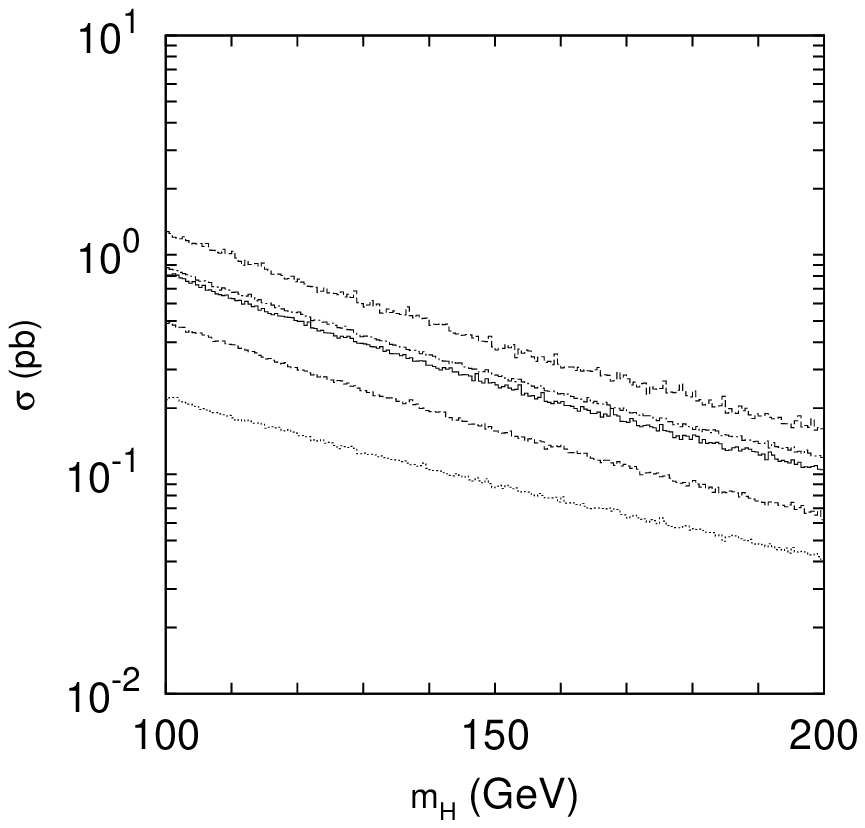, width = 13.0cm}
\caption{The total cross section of associated $b\bar b H$ (upper plot) and $t\bar t H$ (lower plot) 
production as a function of Higgs mass at $\sqrt s = 14$~TeV.
The solid histograms correspond to the results obtained in 
the $k_T$-factorization approach of QCD with the CCFM-evolved unintegrated gluon density (set $A0$).
The upper and lower dashed histograms represent the scale variations
 of $k_T$-factorization predictions, as it was described in the text.
The dash-dotted histograms represent results which were obtained 
in the standard (collinear) approximation of QCD at LO.
The contributions from the quark-antiquark annihilation mechanism 
(multiplied by a factor of $100$ in the case of $b\bar b H$ production)
are shown by the dotted histograms.}
\end{center}
\label{fig3}
\end{figure}

\newpage

\begin{figure}
\begin{center}
\epsfig{figure=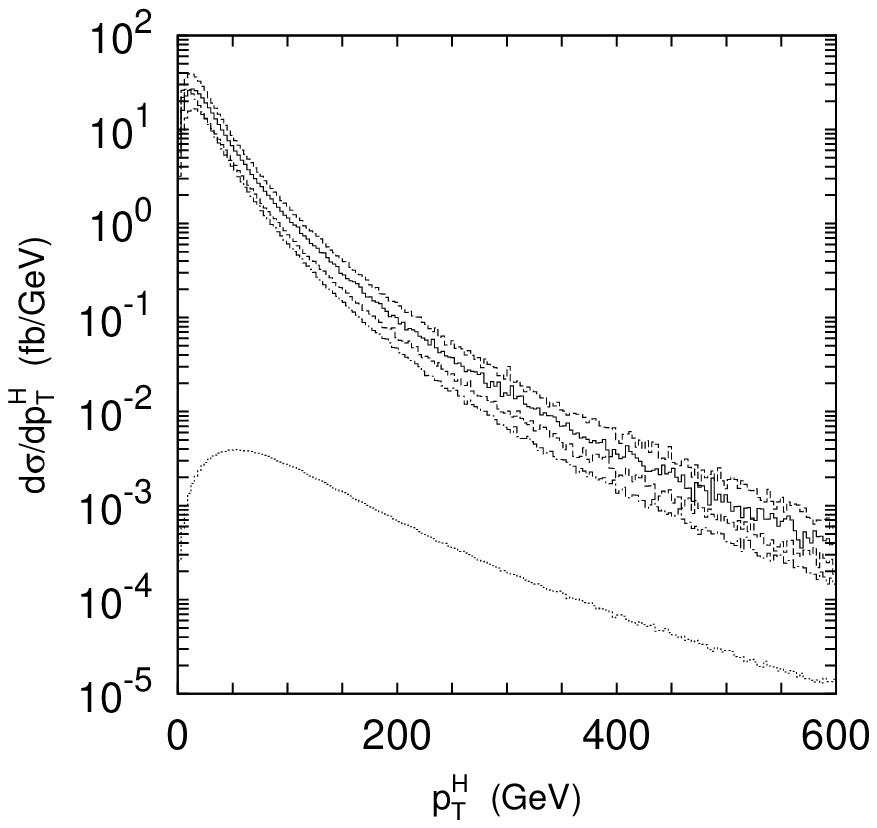, width = 13.0cm}
\epsfig{figure=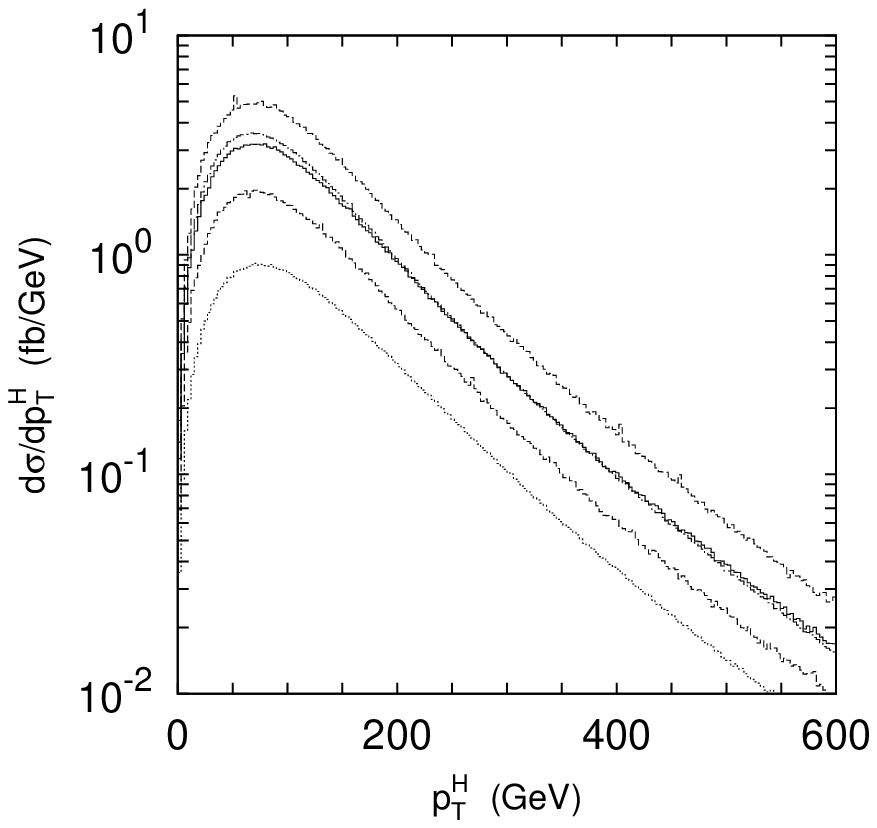, width = 13.0cm}
\caption{The transverse momentum distributions $d\sigma/dp_T^H$ of
associated $b\bar b H$ (upper plot) and $t\bar t H$ (lower plot) 
production calculated at $m_H = 120$~GeV and $\sqrt s = 14$~TeV. 
Notation of the histograms is the same as in Fig.~3.}
\end{center}
\label{fig4}
\end{figure}

\newpage

\begin{figure}
\begin{center}
\epsfig{figure=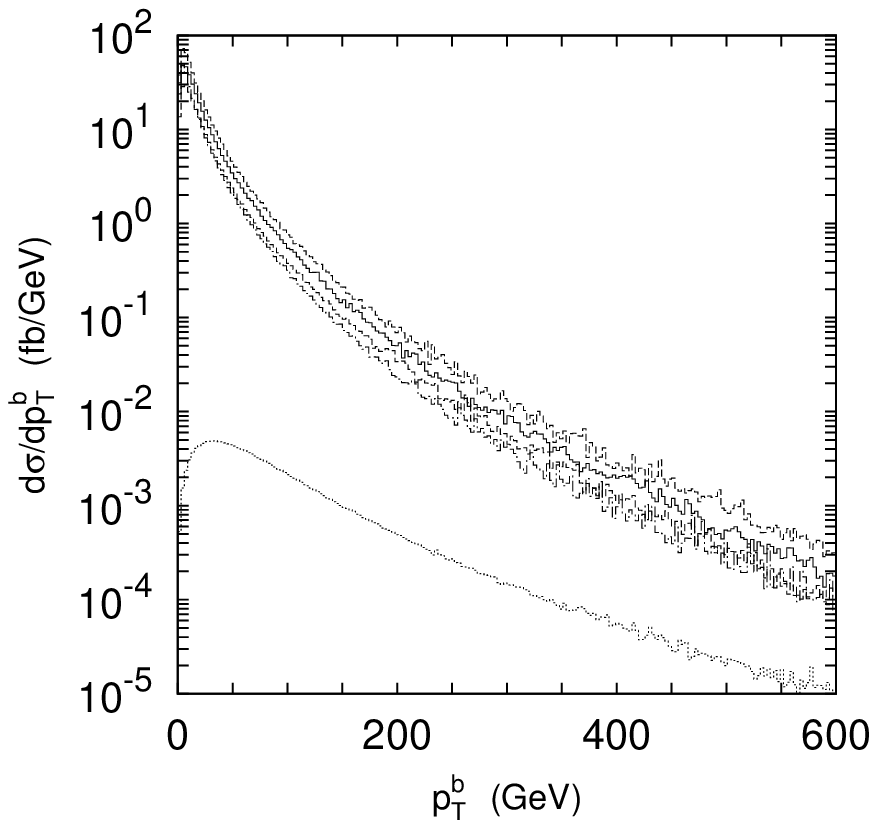, width = 13.0cm}
\epsfig{figure=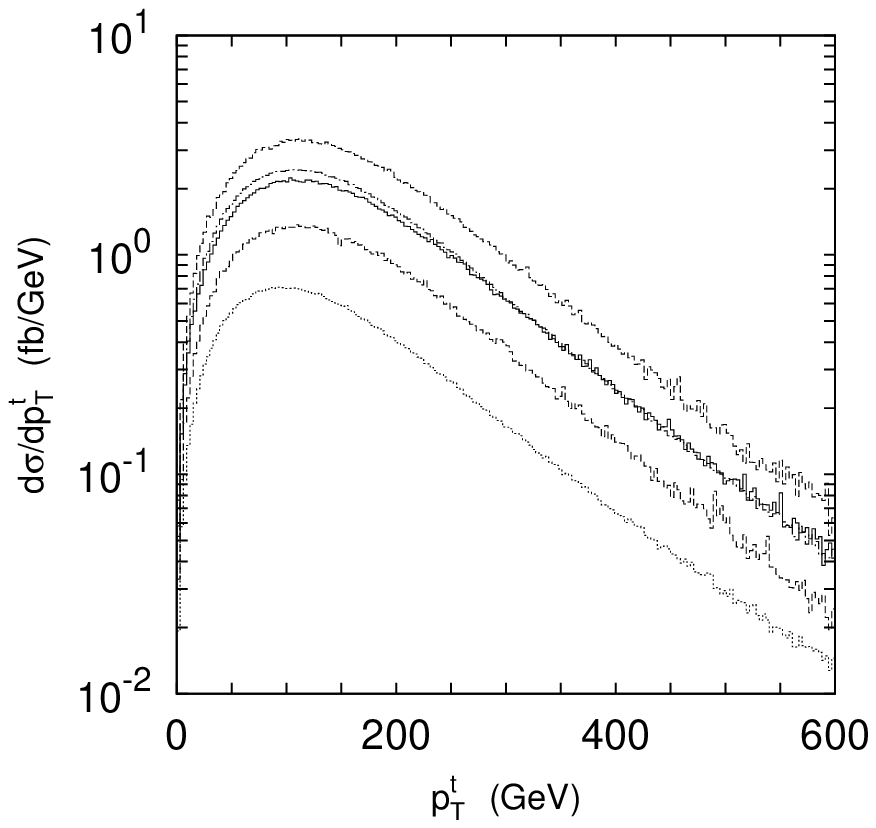, width = 13.0cm}
\caption{The transverse momentum distributions $d\sigma/dp_T^Q$ of
associated $b\bar b H$ (upper plot) and $t\bar t H$ (lower plot) 
production calculated at $m_H = 120$~GeV and $\sqrt s = 14$~TeV. 
Notation of the histograms is the same as in Fig.~3.}
\end{center}
\label{fig5}
\end{figure}

\newpage

\begin{figure}
\begin{center}
\epsfig{figure=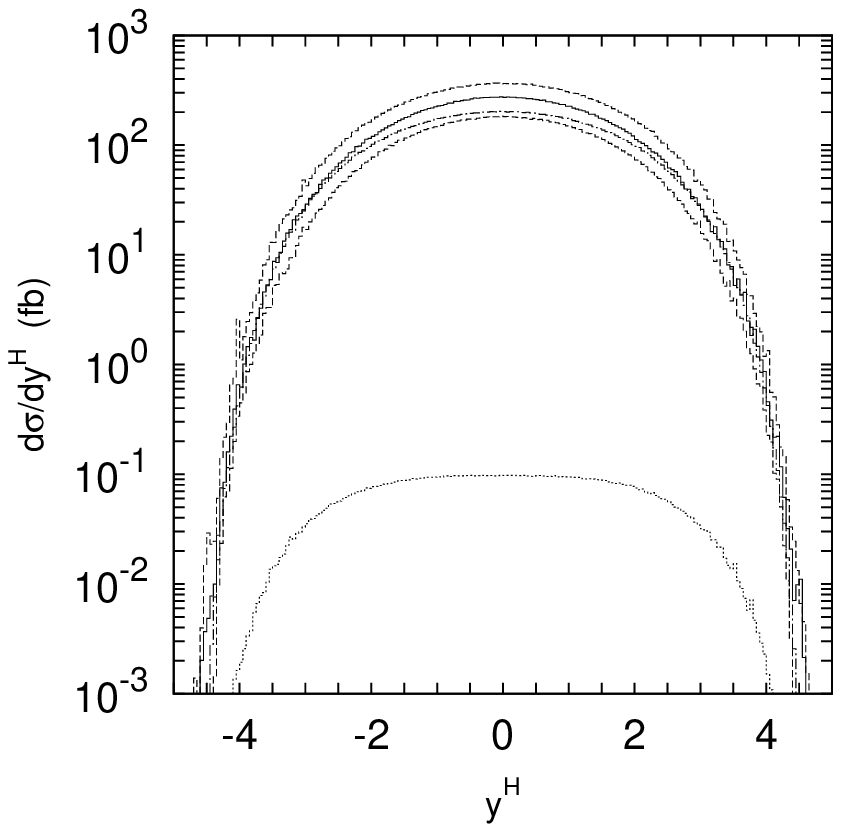, width = 13.0cm}
\epsfig{figure=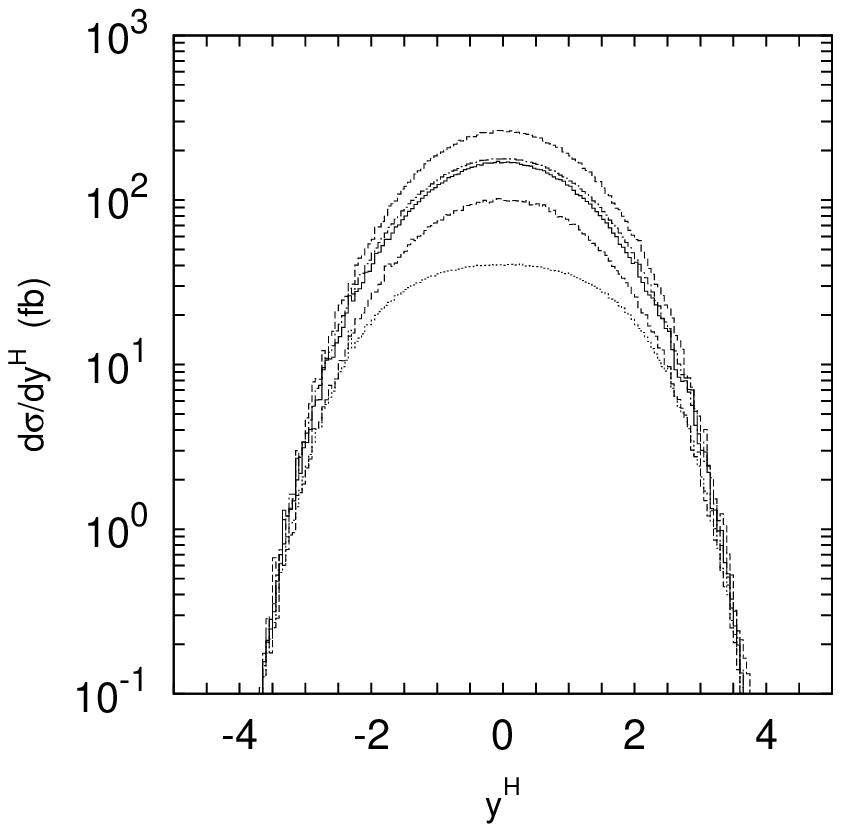, width = 13.0cm}
\caption{The rapidity distributions $d\sigma/dy^H$ of
associated $b\bar b H$ (upper plot) and $t\bar t H$ (lower plot) 
production calculated at $m_H = 120$~GeV and $\sqrt s = 14$~TeV. 
Notation of the histograms is the same as in Fig.~3.}
\end{center}
\label{fig6}
\end{figure}

\newpage

\begin{figure}
\begin{center}
\epsfig{figure=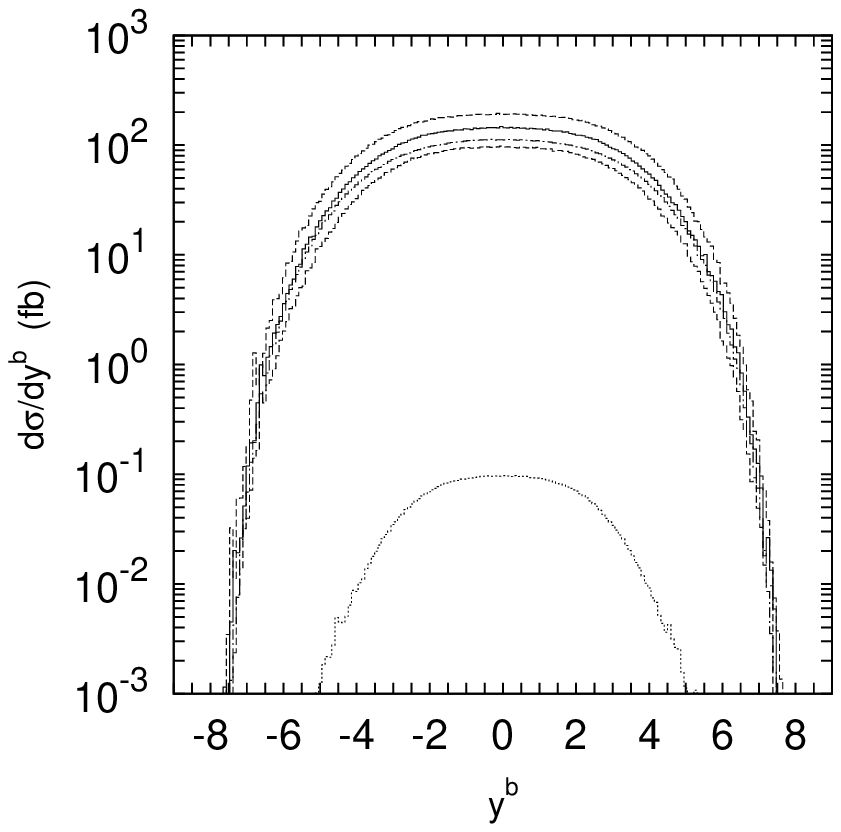, width = 13.0cm}
\epsfig{figure=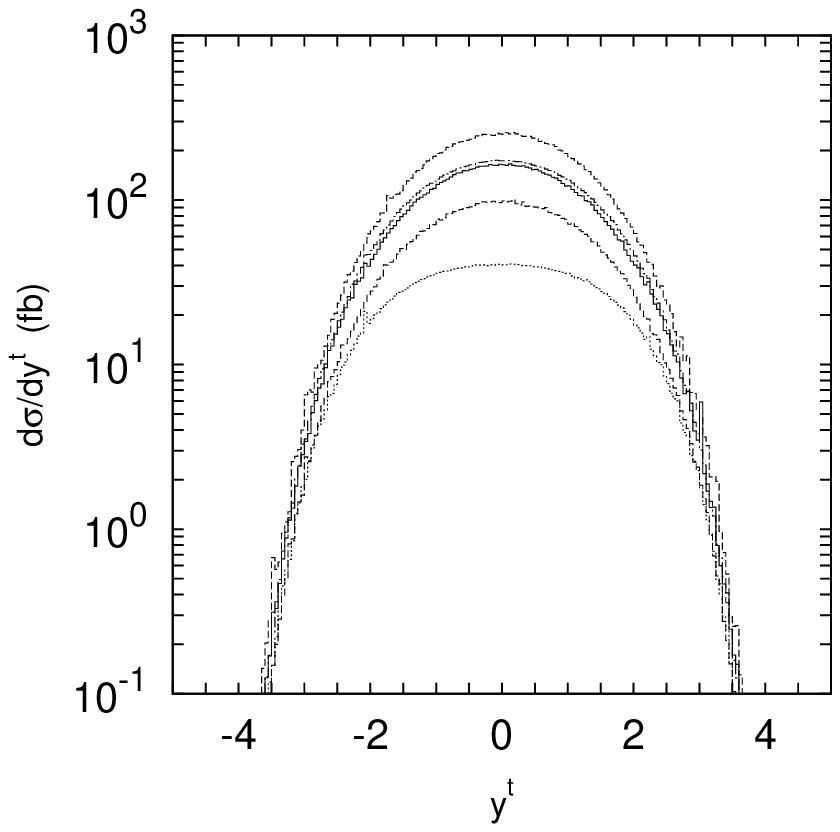, width = 13.0cm}
\caption{The rapidity distributions $d\sigma/dy^Q$ of
associated $b\bar b H$ (upper plot) and $t\bar t H$ (lower plot) 
production calculated at $m_H = 120$~GeV and $\sqrt s = 14$~TeV. 
Notation of the histograms is the same as in Fig.~3.}
\end{center}
\label{fig7}
\end{figure}

\newpage

\begin{figure}
\begin{center}
\epsfig{figure=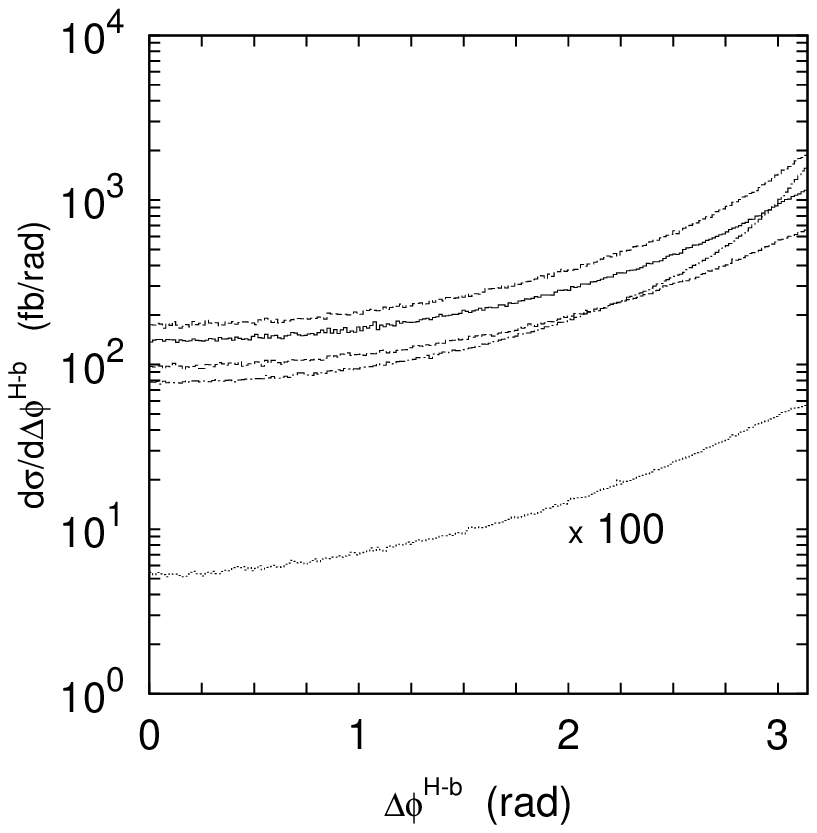, width = 13.0cm}
\epsfig{figure=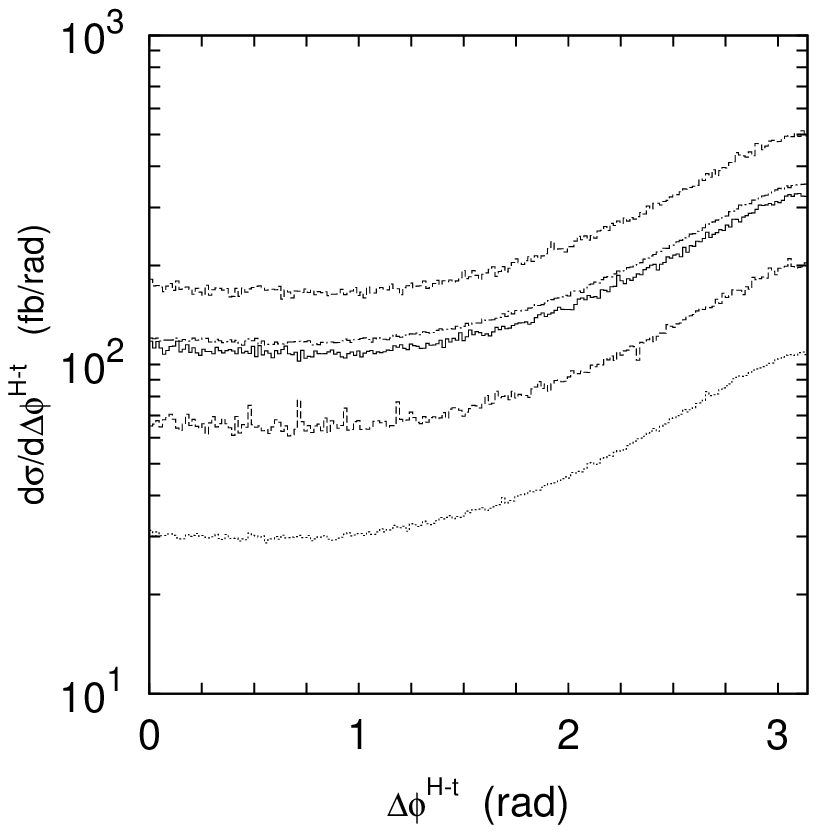, width = 13.0cm}
\caption{The azimuthal angle distributions $d\sigma/d \Delta \phi^{H - Q}$ of
associated $b\bar b H$ (upper plot) and $t\bar t H$ (lower plot) 
production calculated at $m_H = 120$~GeV and $\sqrt s = 14$~TeV. 
Notation of the histograms is the same as in Fig.~3.}
\end{center}
\label{fig8}
\end{figure}

\newpage

\begin{figure}
\begin{center}
\epsfig{figure=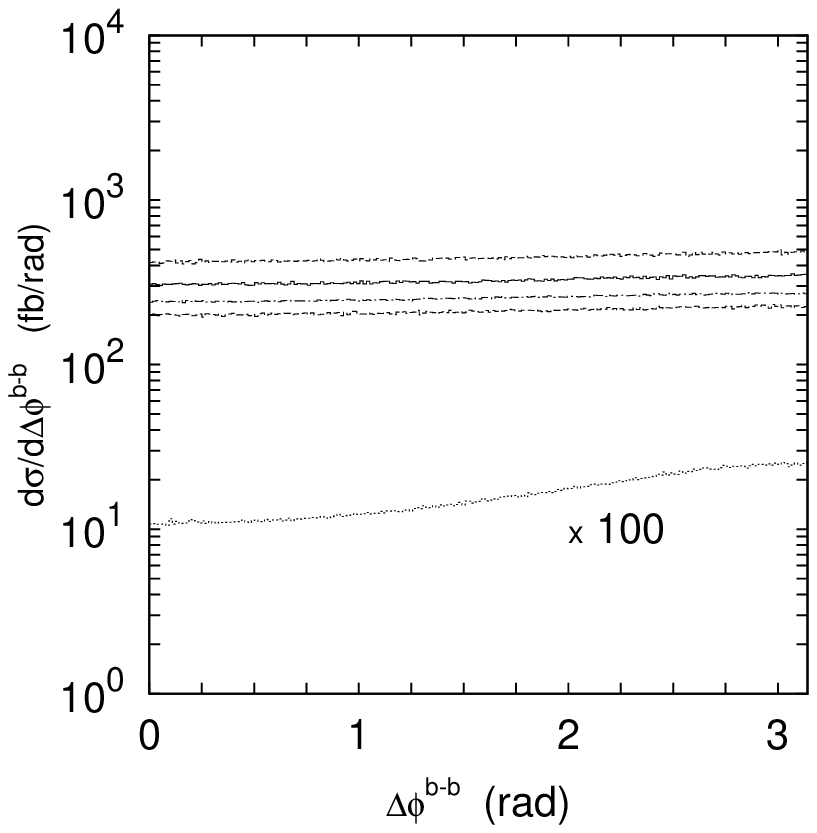, width = 13.0cm}
\epsfig{figure=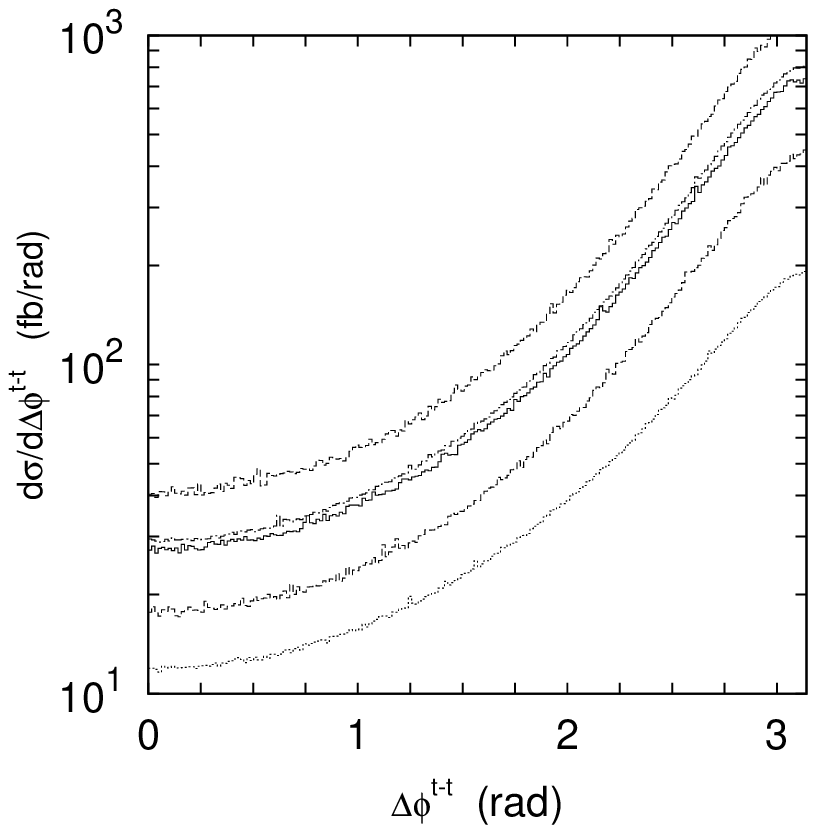, width = 13.0cm}
\caption{The azimuthal angle distributions $d\sigma/d \Delta \phi^{Q - \bar Q}$ of
associated $b\bar b H$ (upper plot) and $t\bar t H$ (lower plot) 
production calculated at $m_H = 120$~GeV and $\sqrt s = 14$~TeV. 
Notation of the histograms is the same as in Fig.~3.}
\end{center}
\label{fig9}
\end{figure}

\newpage

\begin{figure}
\begin{center}
\epsfig{figure=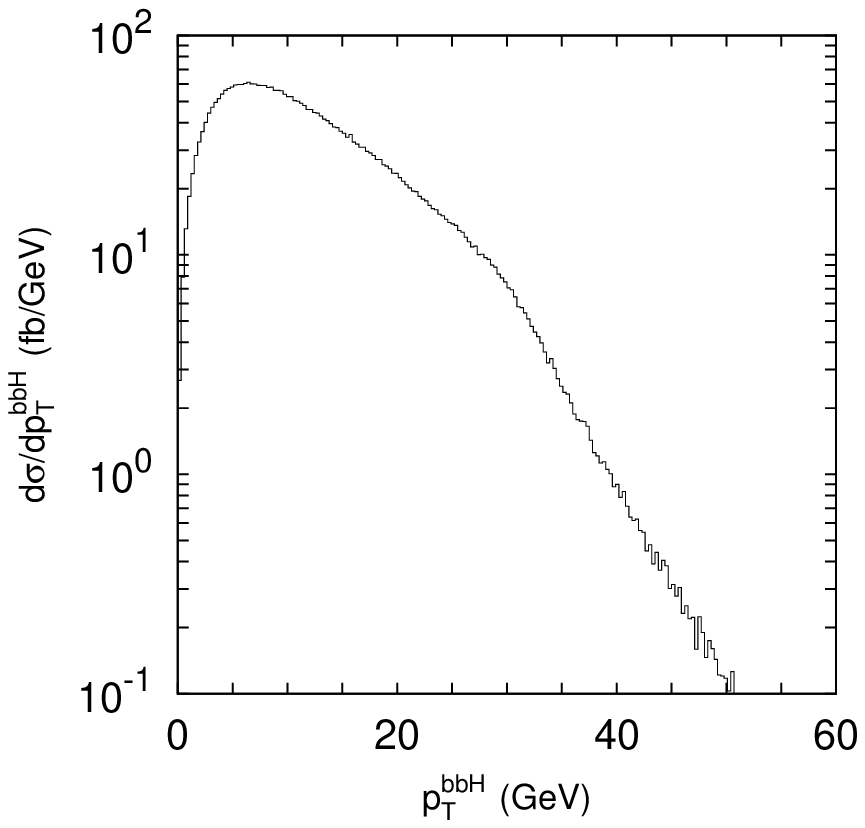, width = 13.0cm}
\epsfig{figure=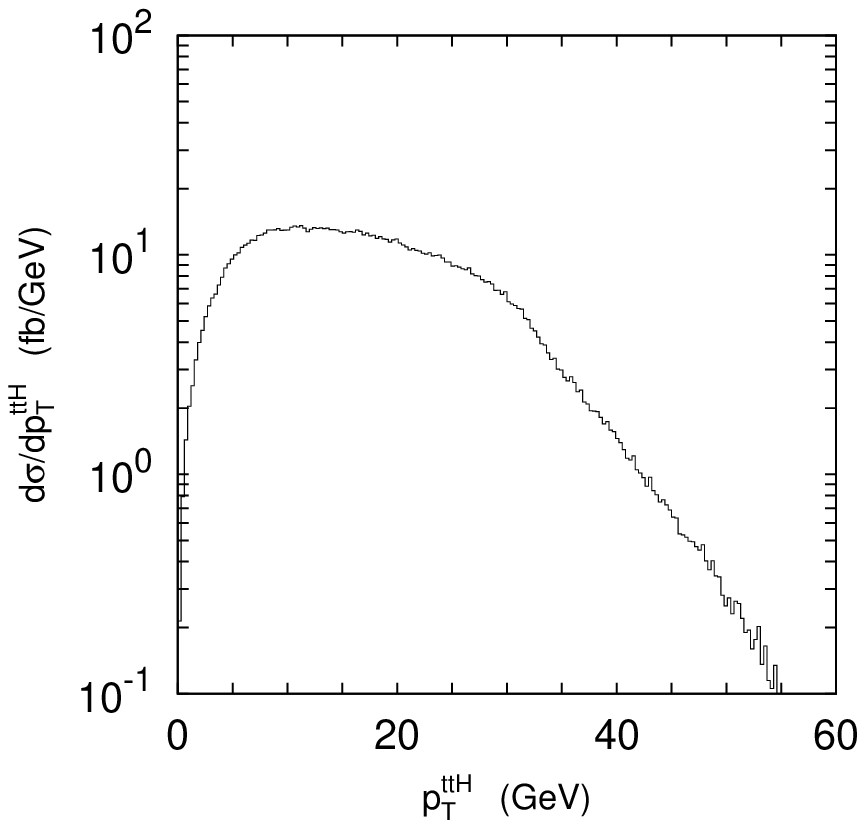, width = 13.0cm}
\caption{The transverse momentum distributions $d\sigma/dp_T^{Q\bar Q H}$ of
associated $b\bar b H$ (upper plot) and $t\bar t H$ (lower plot) 
production calculated at $m_H = 120$~GeV and $\sqrt s = 14$~TeV. 
The off-shell gluon-gluon fusion mechanism is only taken into account.}
\end{center}
\label{fig10}
\end{figure}

\newpage

\begin{figure}
\begin{center}
\epsfig{figure=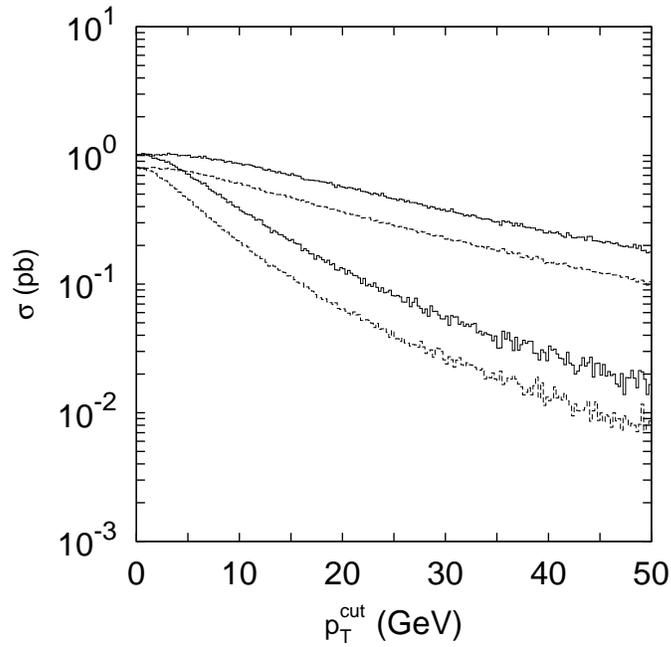, width = 13.0cm}
\caption{The cross sections for ${b\bar b H}$ production 
with one (upper histograms) or two (lower histograms) high-$p_T$ beauty quarks as a function of the
minimal $b$-quark transverse momentum $p_T^{\rm cut}$ calculated at $m_H = 120$~GeV and $\sqrt s = 14$~TeV. 
The solid and dashed histograms correspond to the results obtained in 
the $k_T$-factorization approach and in the collinear LO approximation, respectively.}
\end{center}
\label{fig11}
\end{figure}

\end{document}